\documentclass[10pt,letterpaper,twocolumn]{article} 
\usepackage{ol2_modified}
\usepackage{graphicx} 
\usepackage{amsmath}
\usepackage{bm}  
\usepackage{float}

\newcommand  {\at}     {\mathrm{at}}
\newcommand  {\sub}    {\mathrm{sub}}
\newcommand  {\sat}    {\mathrm{sat}}
\newcommand  {\eff}    {\mathrm{eff}}
\newcommand  {\scat}   {\mathrm{sc}}

\newcommand {\INLN} {Institut Non Lin\'{e}aire de Nice, CNRS and
Universit\'e Nice Sophia-Antipolis,\\ 1361 route des Lucioles, 06560
Valbonne, France}
\newcommand {\Brazil} {CAPES Foundation, Ministry of Education of Brazil, Bras\'{i}lia - DF
70040-020, Brazil}

\begin{document}

\twocolumn[ 

\title{Subradiance in a Large Cloud of Cold Atoms}

\author{William Guerin$^{1,*}$, Michelle O. Ara\'{u}jo$^{1,2}$, Robin Kaiser$^1$}

\affiliation{
$^1$ \INLN \\
$^2$ \Brazil \\
$^*$e-mail: william.guerin@inln.cnrs.fr
}

\begin{abstract}
Since Dicke's seminal paper on coherence in spontaneous radiation by atomic ensembles, superradiance has been extensively studied. Subradiance, on the contrary, has remained elusive, mainly because subradiant states are weakly coupled to the environment and are very sensitive to nonradiative decoherence processes.
Here we report the experimental observation of subradiance in an extended and dilute cold-atom sample containing a large number of particles. We use a far detuned laser to avoid multiple scattering and observe the temporal decay after a sudden switch-off of the laser beam. After the fast decay of most of the fluorescence, we detect a very slow decay, with time constants as long as 100 times the natural lifetime of the excited state of individual atoms. This subradiant time constant scales linearly with the cooperativity parameter, corresponding to the on-resonance optical depth of the sample, and is independent of the laser detuning, as expected from a coupled-dipole model.
\end{abstract}

]


\maketitle


Despite its many applications, ranging from astrophysics~\cite{Baudouin:2014b} to mesoscopic physics~\cite{Labeyrie:2008,Kaiser:2009} and quantum information technology~\cite{Sangouard:2011}, light interacting with a large ensemble of $N$ scatterers still bears many surprising features and is at the focus of intense research. For $N=2$ atoms placed close together, the in-phase oscillation of the induced dipoles produces a large, superradiant dipole, whereas the out-of-phase oscillation corresponds to a subradiant quadrupole. Generalizing for $N\gg 2$, Dicke has shown that, for samples of a small size compared to the wavelength of the atomic transition, the symmetric superposition of atomic states induces superradiant emission, scaling with the number of particles $N$, whereas the antisymmetric superpositions are decoupled from the environment, with vanishing emission rates, which corresponds to subradiance~\cite{Dicke:1954}.

Dicke superradiance has been extensively studied in the 1970s~\cite{Feld:1973,Friedberg:1973,Gross:1982} but the observation of its counterpart, subradiance, has been restricted to indirect evidence of modified decay rates in one particular direction~\cite{Pavolini:1985} or in systems of two particles at very short distance~\cite{DeVoe:1996,Barnes:2005,McGuyer:2015}.
One challenge for the observation of subradiance by a large number of particles is the fragile nature of these states, which require protection from any local nonradiative decay mechanism~\cite{Temnov:2005}.
Furthermore, contrary to the two-atom case, for which the distance between atoms has to be small compared to the wavelength, for $N\gg 2$, the retarded, long-range resonant dipole-dipole interaction~\cite{Stephen:1964} gives rise to super- and subradiant effects (``cooperative scattering'') also in \emph{dilute} samples, with interatomic distances much larger than the wavelength, and corresponding large system sizes. Since, for $N>2$, the Hamiltonians for short and long-range interactions do not commute, the collective eigenstates due to the long-range interactions are suppressed by short-range interactions~\cite{Gross:1982}. These short-range or near-field effects (or `van der Waals dephasing') thus need to be avoided in this case. As a consequence -- and maybe counterintuitively -- a large and dilute sample of interacting dipoles is the most appropriate system for the observation of $N$-body subradiance.

In this regime, and in the weak excitation limit (``single-photon superradiance'')~\cite{Scully:2006,Scully:2009,Roehlsberger:2010}, it has been shown that the superradiant enhancement of the emission rate scales as the cooperativity parameter $N/M$, where $M$ is the number of available modes for the electromagnetic radiation~\cite{Svidzinsky:2008,Svidzinsky:2008b,Oliveira:2014,Longo:2015}. For a spherical sample of radius $R$, $M\sim (k_0 R)^2$, where $k_0=2\pi/\lambda$, this cooperativity parameter is proportional to the peak on-resonant optical depth of the atomic cloud, given by $b_0=3N/(k_0 R)^2$ for a cold-atom cloud with a Gaussian density distribution of rms radius $R$. This number can be large even at low density. In a recent work~\cite{Bienaime:2012}, we used a coupled-dipole model to generalize this result to subradiance (see also the Supplemental Material~\cite{SuppMat}). In this Letter, we report the experimental observation of subradiance in this weak-excitation, dilute- and extended-sample limit.


\begin{figure}[t]
\centering
\includegraphics{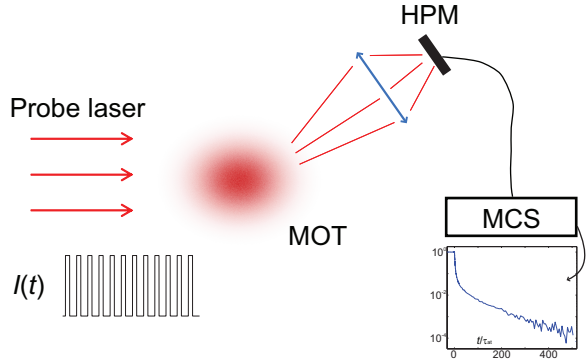}
\caption{(color online). Principle of the experiment. A large probe laser illuminates the atomic sample for $30$~$\mu$s and is switched off rapidly. The fluorescence at $\sim 35^\circ$ is collected by a hybrid photomultiplier (HPM) and recorded on a multichannel scaler (MCS). The experiment is repeated $500 000$ times. At each cycle, 12 pulses are recorded during the free expansion of the cloud, allowing the on-resonance optical depth to vary.}
\label{Fig1}
\end{figure}

In our experiment, we load $N\approx 10^9$ $^{87}$Rb atoms from a background vapor into a magneto-optical trap (MOT) for 50~ms. A compressed MOT (30~ms) period allows for an increased and smooth spatial density with a Gaussian distribution of rms size $R\approx$ 1~mm (typical density $\rho \approx 10^{11}$~cm$^{-3}$) and a reduced temperature $T\approx 50~\mu$K. We then switch off the MOT trapping beams and magnetic field gradient and allow for 3~ms of free expansion, used to optically pump all atoms into the upper hyperfine ground state $F=2$. Next, we apply a series of 12 pulses of a weak probe beam (waist $w = 5.7$~mm), linearly polarized and detuned by $\delta=(\omega-\omega_0)/\Gamma$ from the closed atomic transition $F=2 \rightarrow F'=3$. Here $\omega$ is the frequency of the laser, $\omega_0$ the frequency of the atomic transition (of wavelength $\lambda=2\pi c/\omega_0=780.24$~nm) and $\Gamma/2\pi = 6.07$~MHz its linewidth. Note that when we varied the detuning, we also varied the laser intensity accordingly in order to keep the saturation parameter constant at $s \simeq 4.5\times 10^{-2}$.
The pulses of duration $30~\mu$s and separated by $1$~ms are obtained by using two acousto-optical modulators in series to reach an extinction ratio better than $10^{-4}$. The 90\%-10\% fall time at the switch-off is $\sim 15$~ns, limiting the possibility of studying superradiance, but convenient for detecting subradiance. Between subsequent pulses of each series, the size of the cloud increases because of thermal expansion, and the atom number decreases due to off-resonant optical pumping into the $F=1$ hyperfine state during each pulse. The corresponding change of the on-resonant optical depth $b_0$ allows us to conveniently measure the decay of the fluorescence as a function of $b_0$ and investigate whether $b_0$ is the relevant scaling parameter~\cite{SuppMat}. After this stage of expansion and measurement, the MOT is switched on again and most of the atoms are recaptured. The complete cycle is thus short enough to allow the signal integration over a large number of cycles, typically $\sim 500 000$ (complete acquisition time $\sim 14$h per run). The scattering of the probe beam is collected by a lens with a solid angle of $\sim 5\times 10^{-2}$~sr at  $\theta\approx35^\circ$ from the incident direction of the laser beam (see Fig.~1). We use a hybrid photomultiplier (Hamamatsu HPM R10467U-50) in the photon-counting regime, without any measurable amount of afterpulsing, which would considerably mask signatures of subradiance. The signal is then recorded on a multichannel scaler (MCS6A by FAST ComTec) with a time bin of $1.6$~ns, averaging over the cycles. The cooperativity parameter $b_0$ corresponding to each pulse is calibrated by an independent measurement of the atom number, cloud size and temperature using absorption imaging~\cite{SuppMat}.


\begin{figure*}[t] 
\centering
\includegraphics{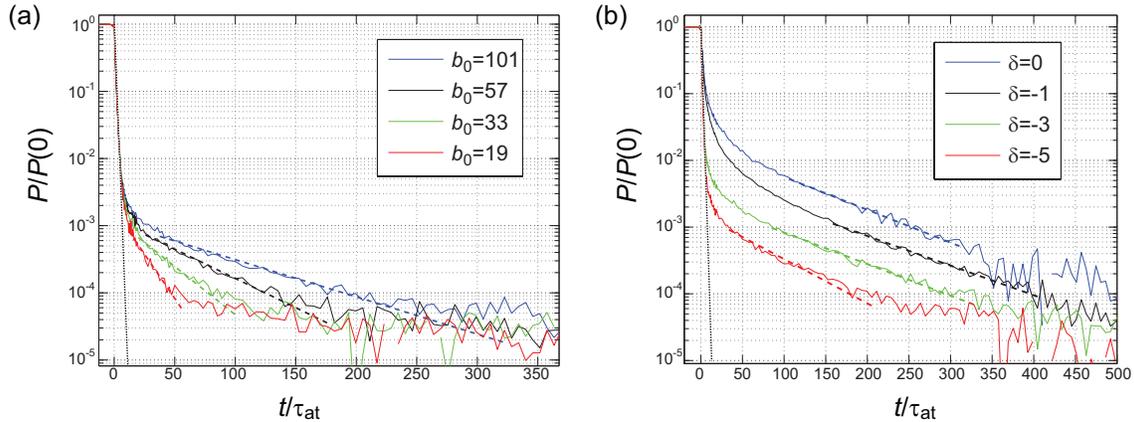}
\caption{(color online). Slow decay of the fluorescence after switching off the probe laser. The vertical scale is normalized by the steady-state detected power $P(0)$ and the time scale is normalized by the atomic lifetime of the excited state $\tau_\at$. Without any collective effect (single-atom physics), the decay would be given by $P(t)=P(0) \exp(-t/\tau_\at)$ (the black dotted line). (a) Several data are shown for different on-resonance optical depths $b_0$ and the same detuning $\delta = -6$ (in units of $\Gamma$). The time constant increases with $b_0$. (b) Several data are shown for different detunings and the same $b_0 = 108 \pm 5$. The time constant remains unchanged, but the relative amplitude of the subradiant decay decreases as the detuning increases.}
\label{Fig2}
\end{figure*}

Typical data are shown in Fig.~2. The signal is normalized to the steady-state fluorescence level and we focus on the switch-off period to highlight the slow fluorescence decay. In Fig.~2(a), the detuning $\delta$ of the probe beam is kept constant and the different decay curves correspond to different values of $b_0$, obtained in a single run. On the contrary, Fig.~2(b) shows data taken with different detunings but for the same $b_0$. In both cases, most of the fluorescence decays fast (note the logarithmic scale of the vertical axis), but a slow decay is clearly seen well above the noise floor (slightly below $10^{-4}$). We stress that fluorescence can be detected at very large delays, as can be seen from the time axis, in units of $\tau_\at = \Gamma^{-1} = 26$~ns. We attribute this slow decay to subradiance in the single-photon (or weak excitation) regime, as predicted in Ref.~\cite{Bienaime:2012}.

A qualitative analysis of the two figures clearly shows different behaviors. As $b_0$ is varied, the slow decay rate changes, whereas its relative amplitude stays approximately the same. The exact opposite happens when we change the detuning, keeping $b_0$ fixed.
For a quantitative analysis, we fit the slow tail at long delays by an exponential decay with two free parameters: the time constant $\tau_\sub$ and its relative amplitude $A_\sub$ \cite{SuppMat}. We systematically studied how these parameters depend on $b_0$ and $\delta$. The result of this analysis is presented in Fig.~3. In Fig. 3(a) we plot the subradiant time constant as a function of $b_0$ for different detunings. The collapse of all points on the same curve clearly indicates that the slow decay rate does not depend on the detuning [see also Fig.~3(b)]. This demonstrates that this slow decay is not a multiple-scattering effect, such as previously observed radiation trapping~\cite{Labeyrie:2003}, which depends on the optical depth at the laser frequency $b(\delta) \propto b_0/(1+4\delta^2)$, with a strong dependance with $\delta$.
The second feature in  Fig.~3(a) is the linear increase of $\tau_\sub$ with $b_0$, up to time constants as long as $\tau_\sub \sim 100 \tau_\at$. This is perfectly consistent with the predictions of the coupled-dipole model for subradiance \cite{SuppMat}. We note that for large negative detunings, one has to take into account the variation of $b_0$ during the pulse series induced by the cloud expansion together with a significant contribution of atom losses by off-resonant hyperfine pumping. This allowed us to test different combinations of $N$ and $R$ as scaling parameters (see Fig.~\ref{figS2} in the Supplemental Material \cite{SuppMat}). The comparison showed that the best collapse has been obtained with $N/R^2 \propto b_0$ as the scaling parameter, which demonstrates that $b_0$ is indeed the relevant cooperativity parameter in the regime of a dilute and extended sample, as expected from the ratio of the number of atoms to the number of available electromagnetic modes radiating from the sample. Finally, we show in Fig.~3(c) the relative amplitude $A_\sub$ of the slow decay. As was already seen in Fig.~2(b), this amplitude is much larger near resonance, and seems to reach a plateau for large detunings. We have checked that subradiant decay is still visible at a larger detuning, up to $\delta=-11$~\cite{SuppMat}. This is in line with the coupled-dipole model, in which the weight of the long-lived modes are enhanced near resonance and the weight of all of the collective modes becomes independent of the detuning at large detuning.

As a long lifetime can also occur due to multiple scattering, when the optical depth $b(\delta) \gg 1$~\cite{Labeyrie:2003}, we investigated the decay time close to the atomic resonance -- to study how subradiance compares with radiation trapping. In the range of the accessible experimental values [Fig.~3(b)], no marked difference of the decay times around resonance is visible, even though a small difference was predicted in Ref.~\cite{Bienaime:2012}. The interplay of radiation trapping and subradiance near resonance is still an open question and will be the subject of our further experiments.

\begin{figure*}[t] 
\centering
\includegraphics{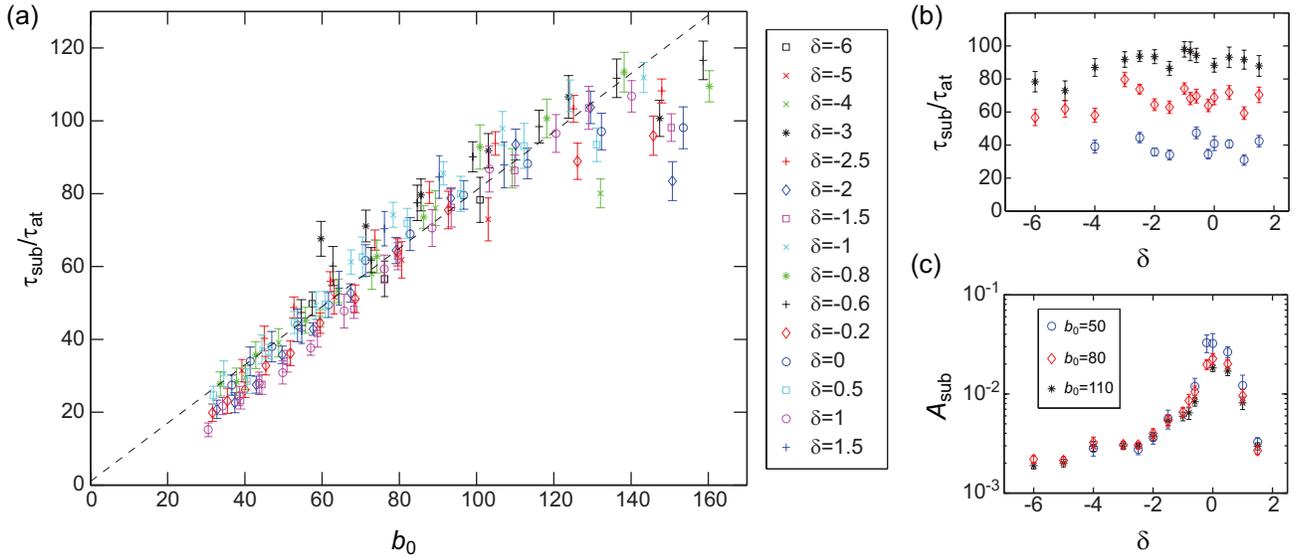}
\caption{Dependence of the subradiance time with the on-resonance optical depth $b_0$ and the laser detuning $\delta$. (a) Subradiant time constant $\tau_\sub$ (in units of $\tau_\at$) as a function of $b_0 = 3N/(k_0R)^2$ for different detunings of the probe laser. Almost all of the points collapse on a single curve, showing the linear scaling with $b_0$. The dashed line is a linear fit $\tau_\sub/\tau_\at = 1 + \alpha b_0$ with the slope $\alpha$ as a free parameter. Excluding the data with $b_0>120$ we obtain $\alpha=0.8$. (b) From the same data, $\tau_\sub$ is plotted as a function of the detuning for three different values of $b_0$, illustrating that $\tau_\sub$ is independent of the detuning. (c) Similarly, the subradiant relative amplitude $A_\sub$  is plotted as a function of the detuning for the same three different values of $b_0$. Subradiance is more important near resonance and decreases towards a plateau at large detuning.}
\label{Fig3}
\end{figure*}

We have also studied the effect of the probe intensity and checked that, at low saturation parameter, the observed subradiance is independent of the intensity (see Fig.~\ref{FigS3} in the Supplemental Material~\cite{SuppMat}), which validates the use of the coupled-dipole model in the weak-excitation limit. We finally also excluded the possibility that residual near-resonant light might always be present and induce a slow decay due to radiation trapping, thus mimicking off-resonant subradiance~\cite{SuppMat}.


To summarize, we have presented the first direct signatures of subradiance in a large system of resonant scatterers. We have shown that in the regime of dilute and extended samples, the subradiant decay rate is governed by a cooperativity parameter defined as the ratio of the number of scatterers and the sample size squared, which conveniently corresponds to the on-resonance optical depth. This observation of subradiance opens interesting questions, including the robustness of subradiance against decoherence mechanisms or the possibility of controlling the population of the subradiant modes by an appropriate temporal or spatial shaping of the driving laser or of the atomic levels. If the subradiant states can be manipulated with sufficient control~\cite{Scully:2015}, their isolation from the environment might be exploited as a resource for quantum information or quantum metrology~\cite{Ostermann:2013}. As subradiance goes hand in hand with superradiance, simultaneous recording of fast and slow decays would be a beautiful illustration of the cooperative scattering envisioned by Dicke. By using a stronger laser drive, it would also be possible to access a larger subspace of the full Hilbert space, addressing the possibility of a photon-blockade effect~\cite{Pellegrino:2014}.

In addition to quantum optics, our observation is also relevant to mesoscopic physics~\cite{Rossum:1999,AkkermansMontambaux}, a community less familiar with Dicke physics. One major challenge in this field is the observation of strong localization of light, the analogy for classical waves of Anderson localization of electrons~\cite{Anderson:1958}. Previous experimental observations of a decay of scattered light slower than predicted by the diffusion equation have been used as a signature of Anderson localization in dielectrics~\cite{Maret:2006,Scheffold:2013}. Our results show that it cannot always be the case with cold atoms. Similarly, recent numerical simulations considering point-dipole resonant scatterers study the collective modes of the effective Hamiltonian of the system and, in particular, their lifetimes~\cite{Akkermans:2008,Skipetrov:2014,Bellando:2014,Skipetrov:2015,Maximo:2015}. Our work shows that Dicke subradiance can also be at the origin of very long lifetimes and that a careful analysis is required to distinguish subradiant from localized modes~\cite{Maximo:2015}. Finally, the combination of subradiance with disorder acting on the atomic transitions might provide an alternative route to strong localization of light, as was recently suggested~\cite{Biella:2013}.

We acknowledge fruitful discussions with Nicola Piovella, Tom Bienaim\'e, Romain Bachelard, Guillaume Labeyrie, and Dominique Delande and the technical help from Louis Bellando, Ivor Kresic, Lo\"ic Lavenu, and Antoine Dussaux, and we thank Alain Aspect for his constructive comments on the manuscript. We also acknowledge financial support from the French Agence National pour la Recherche (project LOVE, Grant No. ANR-14-CE26-0032), the Brazilian Coordena\c{c}\~{a}o de Aperfei\c{c}oamento de Pessoal de N\'{i}vel Superior (CAPES), the Brazilian Conselho Nacional de Desenvolvimento Cient\'ifico e Tecnol\'ogico (project PVE, Grant No. 303426/2014-4), and the European Research Executive Agency (program COSCALI, Grant No. PIRSES-GA-2010-268717).





\twocolumn[

\LARGE

\centerline{\textbf{Supplemental Material}}

\ \\

\normalsize

]

\section{Experimental details}

\subsection{Calibration of $\mathbf{b_0}$.}

As we cannot measure $b_0$ simultaneously to the data acquisition, the calibration is done in a separate time sequence, using absorption imaging, requiring a longer cycle without recapture of the atoms. We therefore increase the loading time of the MOT such that the fluorescence of the probe beam measured at large detuning is the same as in the measurement cycle, all other parameters being unchanged. The absorption images then provide the transverse profile of the probe transmission $T$, which is related to the optical depth by $T(x,y) = \exp[-b'(x,y)]$, where
\begin{equation}
b'(\delta) = \mathcal{C} \frac{b_0}{1+4\delta^2}\; ,
\end{equation}
with $\delta$ the probe detuning and $\mathcal{C}=7/15$ is the average Clebsch-Gordan coefficient of the transition for a statistical mixture of the Zeeman substates. We use a Gaussian fit for $b'(x,y)$ and extract the r.m.s. size $R$ directly from the fit, as well as $b_0$ from the amplitude of the Gaussian fit and the atom number $N$ from its integral.
This is done for different values for the time of flight of ballistic expansion, from which we extract the temperature, allowing us to know the size of the atomic cloud for the different pulses as well as the initial atom number.

During the pulse series, we need to take into account possible optical pumping by the probe beam into the other hyperfine ground state $F=1$. This effect is almost negligible near resonance but becomes significant for the largest negative detunings used in this work. We therefore have precisely measured the pumping rate, from the fluorescence decay, at very large negative detuning ($\delta = -9$), and used this rate to calibrate the probe intensity, which is measured simultaneously to the data on a separate photodetector. Then we use this intensity to compute the atom losses induced by optical pumping and correct the atom number for each pulse.

Note that the uncertainty on the precise value $\mathcal{C}$ corresponding to the experimental conditions introduces a systematic uncertainty on $N$ and $b_0$, which is thus also affecting the precise value of the slope extracted from the data of Fig.~3. Statistical uncertainties, estimated by the shot-to-shot fluctuations of $b_0$, are of the order of $8\%$ (standard deviation).

\subsection{Analysis of the fluorescence decay.}

The raw data are histograms of number of photons detected with a time bin of 1.6~ns. Our detection in the photon counting regime yields about $10^7$ counts per second (cps) during the pulses, while between pulses, after the end of the subradiant decay, the `dark' level is about $4000$~cps, mainly due to stray light. For each pulse we subtract this dark level and normalize the signal to the steady state amplitude. Then we further bin the data to improve the signal to noise ratio. We use a variable binning, i.e., a larger bin size for later time, when the signal varies slowly, than for short times. We also implemented a systematic procedure to choose the range of data to be fitted by selecting data one decade above the noise floor. This allows us to fit the last measurable time constant. We found this procedure to give the most reliable and relevant fit. Finally, we compute the statistical coefficient of determination of the fit, which quantifies its relevance, and we keep only the points for which this coefficient is above $0.97$ in Fig.~3 of the main paper.

The data gathered in this figure have been taken in exactly the same conditions (initial atom number, temperature, saturation parameter, experimental cycle), where the only changing parameters are the probe detuning and its intensity to keep the saturation parameter constant. We have also been able to observe subradiant decay for larger detunings, up to $\delta=-11$, but we have not included these data because we had to adapt other parameters. Indeed, because of the limited available power for the probe beam, the saturation parameter and the subsequent amount of fluorescence were lower, which we partially compensated by increasing the atom number via the loading time, which changed also the temperature and the cycle duration.

\section{Predictions of the coupled-dipole model}

We recall here the main ingredients of the coupled-dipole model, which has been widely used in the last years in the context of single-photon superradiance \cite{Scully:2006_,Svidzinsky:2008_,Svidzinsky:2008b_,Courteille:2010_,Svidzinsky:2010_,Bienaime:2011_,Bienaime:2012_,Bienaime:2013_,Miroshnychenko:2013_,Feng:2014_}.
We consider $N$ two-level atoms (positions $\bm{r}_i$, transition wavelength $\lambda = 2\pi/k_0$, excited state lifetime $\Gamma^{-1}$) driven by an incident laser (Rabi frequency $\Omega$, detuning $\Delta$, plane wave of wavevector $\bm{k}_0$). Restricting the Hilbert space to the subspace spanned by the ground state of the atoms $|G\rangle = |g \cdots g \rangle$ and the singly-excited states $|i\rangle = |g \cdots e_i \cdots g\rangle$ and tracing over the photon degrees of freedom, one obtains an effective Hamiltonian describing the time evolution of the atomic wave function $| \psi(t) \rangle$,
\begin{equation}
| \psi(t) \rangle = \alpha(t) | G \rangle +  \sum\limits_{i=1}^N \beta_{i}(t)| i \rangle \; . \label{eq.psi}
\end{equation}
Using standard approximations, the effective Hamiltonian can be written as
\begin{equation}
\begin{split}
H_\eff = \frac{\hbar\Omega}{2}\sum_i \left[ e^{i\Delta t -i \bm{k}_0\cdot \bm{r}_i}S_-^i + e^{-i\Delta t +i \bm{k}_0\cdot \bm{r}_i}S_+^i\right] \\
 - \frac{i\hbar \Gamma}{2} \sum_i S_+^i S_-^i -\frac{\hbar\Gamma}{2} \sum_i \sum_{j\neq i} V_{ij} S_+^i S_-^j \;, \end{split}
\label{eq.Heff}
\end{equation}
where $S_\pm^i$ and $S_z^i$ are the usual pseudospin operators for the kets $g_i$ and $e_i$, respectively. The first term of $H_\eff$ describes the coupling to the laser field, the second accounts for the finite lifetime of the excited states and the third one describes the dipole-dipole interactions, with
\begin{equation}
V_{ij} = \frac{e^{ikr_{ij}}}{kr_{ij}} \; , \quad r_{ij} = |\bm{r}_i - \bm{r}_j| \; . \label{eq.Vij}
\end{equation}
Here, we have considered a scalar model for light, which neglects polarization effects and near-field terms in the dipole-dipole interaction. It is known to be a good approximation for dilute clouds \cite{Skipetrov:2014_,Bellando:2014_}, i.e., when the typical distance between atoms is much larger than the wavelength, which is the case in the experiment.

Considering the low intensity limit, when atoms are mainly in their ground states, i.e., $\alpha \simeq 1$, the problem amounts to determine the amplitudes $\beta_i$, which are then given by the linear system of coupled equations
\begin{equation}
\dot{\beta}_i = \left( i\Delta-\frac{\Gamma}{2} \right)\beta_i -\dfrac{i\Omega}{2}e^{i\bm{k}_0 \cdot \bm{r}_i } + \frac{i\Gamma}{2} \sum_{i \neq j} V_{ij}\beta_j \; .
\label{eq.betas}
\end{equation}
These equations are the same as those describing $N$ classical dipoles driven by an oscillating electric field \cite{Svidzinsky:2010_}, justifying the term ``coupled-dipole model''. The first term corresponds to the natural evolution of the dipoles (oscillation and damping), the second one to the driving by the external laser, the last term corresponds to the dipole-dipole interaction and is responsible for all collective effects, including dephasing and attenuation of the driving laser beam, as well as more subtle multiple scattering and cooperative effects (super- and sub-radiance). Note that even if the detuning does not appear explicitly in the dipole-dipole interaction term, it still strongly influences the collective behaviour of the system through the population of the eigenmodes that contribute to the system response to the driving field. At large detuning for instance, multiple scattering vanishes.

From the computed values of $\beta_i$, we can derive the intensity of the light radiated by the cloud as a function of time and of the angle \cite{Bienaime:2011_}. The time dependence of the total radiated power $P$ after switching off the laser is proportional to the derivative of the total excited state population,
\begin{equation}
P \propto -\frac{d}{dt} \sum_{j=1}^N \left| \beta_i(t) \right|^2 \; .
\label{eq.Ptot}
\end{equation}

We have used this model to study superradiance and subradiance decay in more detail than in the previous work~\cite{Bienaime:2012_}. The complete study will be published elsewhere~\cite{Araujo:tobepublished_}. Here, we only show the most important result for the subradiance experiment, which is the linear scaling of the decay time with the parameter $b_0 = 3N/(k_0R)^2$, which also corresponds to the on-resonance optical depth through the center of the cloud. Here $N$ is the atom number and $R$ the r.m.s. radius of the atomic Gaussian density distribution, which are the two parameters of the simulation. However, in order to determine if $b_0$ is the scaling parameters, and not, e.g., the atomic density, we need to change $b_0$ while keeping the density $\rho \propto N/R^3$ constant, by changing simultaneously $N$ and $R$. Since the atom number $N$ is limited to a few thousands in our numerical simulations, quite large densities are required to reach large $b_0$. It is then important to use an exclusion volume when drawing the random positions of the atoms in order to exclude close pairs of atoms~\cite{Bellando:2014_}. We set our exclusion volume as $k_0 r_{ij} > 3$.

\begin{figure}[t]
\centering
\includegraphics{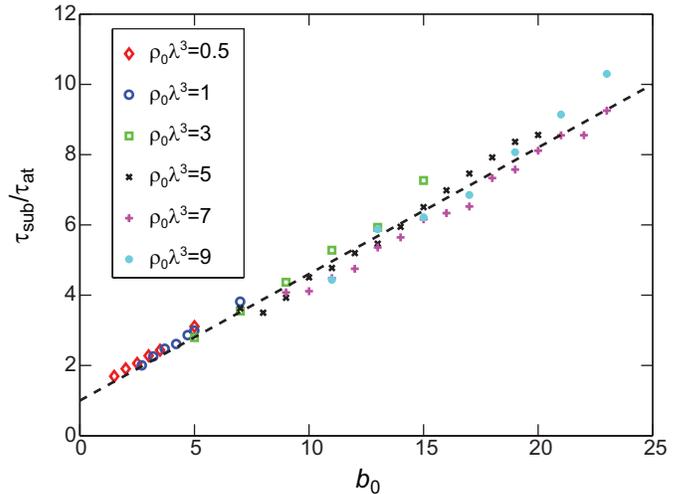}
\caption{Scaling of the the subradiant decay time constant as a function of $b_0$ in the coupled dipole model. We used different densities $\rho_0\lambda^3 = \{0.5, 1, 3, 5, 7, 9\}$. The time constant is extracted from the exponential fit of the end of the decay of the total emitted power computed from Eqs.~(\ref{eq.betas}-\ref{eq.Ptot}). The fit window is chosen for $P$ (normalized to 1 at $t=0$) to be between $10^{-4}$ and $10^{-6}$.}
\label{FigS1}
\end{figure}

\begin{figure*}[t]
\centering
\includegraphics{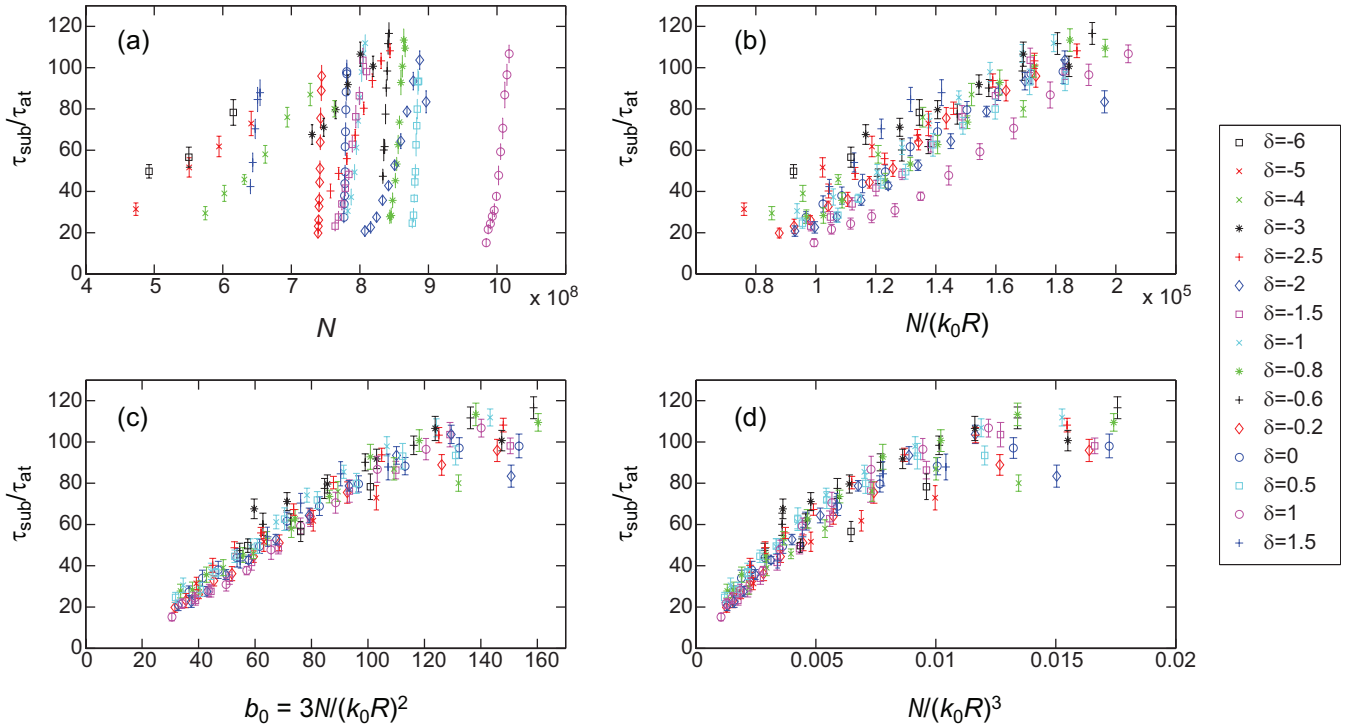}
\caption{Test of different scaling parameters. The subradiant decay time constant is plotted as a function of different combinations of $N$ and $R$. The collapsing of all points on a single curve is best with $b_0$ as the scaling parameter.}
\label{FigS2}
\end{figure*}

In Fig.~\ref{FigS1} we show the result of such simulations, for which we varied $b_0$ at constant density, for different values of $\rho_0\lambda^3 = (2\pi)^{3/2} N/(k_0R)^3$, where $\rho_0$ is the peak density (at the center of the Gaussian cloud). We clearly see that the different curves collapse on the same line, showing that the on-resonance optical depth is indeed the parameter that controls the subradiant decay. The slope of the line is $0.36$. Note that we do not \emph{a priori} expect this number to be in quantitative agreement with the experimental results, as polarization effects and the complex Zeeman structure of the rubidium atoms used in our experiment are neglected in the model.

\section{Test of different scaling parameters}

During the series of pulses, two effects contribute to a change of $b_0$. The most important one is the expansion of the cloud due to its thermal velocity distribution. Its r.m.s. size typically varies from 0.5 to 1.2~mm between the first and the last pulse. This variation is independent of the probe laser detuning. Another contribution to the change of $b_0$ between subsequent pulses is the loss of atoms due to optical pumping into the $F=1$ ground state. Since we kept the same saturation parameter for all detuning, the number of scattered photons is always the same, about 25 per atom and per pulse. However, the probability that a scattering event transfers the atom to the $F=1$ state depends strongly on the detuning from the $F=2 \rightarrow F'=2$ transition, and thus it is much larger for large red detuning $\delta$ (defined from the $F=2 \rightarrow F'=3$ transition). For example, at $\delta=-6$, the number of atoms in the $F=2$ ground state has decreased by 70\% between the first and the last pulse, whereas on resonance it is almost constant for all pulses. Thus, for the different detunings the relative contributions of the variation of $N$ and $R$ are different.

We have exploited this fact to test which combination of $N$ and $R$ allows the different curves acquired with different detunings to collapse on a single curve. This is shown in Fig.~\ref{FigS2}, where it clearly appears that the combination $N/(k_0R)^2$ provides the best scaling parameter. This experimentally confirms that $b_0$ is indeed the parameter governing subradiance in dilute and extended samples, as predicted by the coupled-dipole model (Fig.~\ref{FigS1}).

\section{Subradiance as a function of the probe intensity}

\begin{figure*}[t]
\centering
\includegraphics{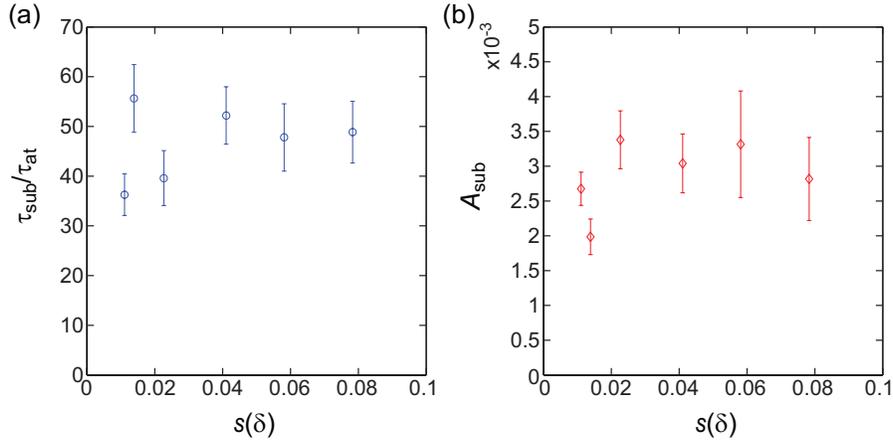}
\caption{Effect of the probe intensity. Subradiance time constant (a) and relative amplitude (b) as a function of the saturation parameter. The data have been taken with $b_0=110\pm 8\%$ and $\delta = -5$.}
\label{FigS3}
\end{figure*}

In order to validate the weak excitation assumption, we have also studied the subradiance decay as a function of the probe intensity. In the linear optics regime, the probe intensity should not impact on the measured decay rates or normalized amplitudes. We have thus fixed the detuning $\delta$ and the optical depth and varied the intensity $I$. On Fig.~\ref{FigS3} we plot the results for the subradiant time constant and relative amplitude as a function of the saturation parameter
\begin{equation}
s(\delta) = \frac{I/I_\sat}{1+4\delta^2} \; ,\label{eq.sat_param}
\end{equation}
where $I_\sat$ is the saturation intensity of the transition.

No significant dependence on the saturation parameter in the explored range ($10^{-2} < s(\delta) < 10^{-1}$) is observed, confirming that the other experimental data, for which $s(\delta)\simeq 0.05$, have been taken in the linear regime.

\section{Possible role of radiation trapping of near-resonant light}

The very slow decays observed at large detuning cannot be simply interpreted as radiation trapping, as in the experiments by Labeyrie {\it et al.} \cite{Labeyrie:2003_,Labeyrie:2005_}, since radiation trapping does not depend on $b_0$ only but depends on the optical depth `seen' by the laser beam, i.e., $b'(\delta) \propto b_0 /(1+4\delta^2)$, which is very small at large detuning.

However, since the amplitude of the subradiance decay is very small, we have to insure that it cannot be explained by radiation trapping of a small amount of light that would \emph{always} be on resonance.

We have investigated two possible nonnegligible sources of near-resonant light: inelastic scattering due to the atomic saturation and the wings of the laser spectrum.

\subsection{Inelastic scattering}

The scattering rate $\Gamma_\scat$ of an atom driven by a laser is the sum of two contributions, elastic and inelastic scattering, given by \cite{Cohen:PetitLivreRouge-EN_}
\begin{equation}
\Gamma_\scat = \Gamma_\mathrm{el} + \Gamma_\mathrm{inel} = \frac{\Gamma}{2} \left( \frac{s}{(1+s)^2} + \frac{s^2}{(1+s)^2}\right) = \frac{\Gamma}{2} \, \frac{s}{1+s} \; ,
\label{eq.scatteringRate}
\end{equation}
where $s$ is the saturation parameter defined above [Eq.~(\ref{eq.sat_param})]. Usually, inelastic scattering is neglected when $s \ll 1$, which is the case in our experiment. However since subradiance is a small signal, special care is required. For $s\ll 1$,
the \emph{relative} weight of inelastic scattering compared to the total scattering is equal to $s$. The spectrum of this inelastic scattering is the well-known Mollow triplet~\cite{Mollow:1969_}, which is, in fact, a triplet only at large $s$. At large detuning and moderate intensity, it is a doublet constituted of two lines of equal weights, one of which is precisely on the atomic resonance~\cite{ChanelierePhD_}. Thus, we expect that after the first scattering event, a proportion $s/2$ of the light is shifted near resonance. Multiple scattering could then trap this light in the sample.

To estimate the weight of this inelastic component in the slow decay, we multiply this $s/2$ fraction by the relative amplitude of the slow decay for resonant photon, which is what we extract from our experiment when the laser is tuned to resonance ($A_\sub \simeq 2\times 10^{-2}$ for $b_0 = 110$, see Fig.~3c of the main paper). In this scenario, we would thus expect $A_\sub \simeq s\times 10^{-2}\ll 10^{-2}$. On the contrary, we see on Fig.~\ref{FigS3} (also with $b_0 = 110$) that $A_\sub$ is significantly larger and in any case not proportional to $s$. These data allow us to conclude that spurious resonant radiation trapping of one Mollow sideband is much smaller than the subradiance we observe.

Note that in this experiment, a laser beam larger than the atomic cloud has been used, leading to a significant proportion of single and low-order scattering events on the edges of the cloud, even on resonance, contrary to the previous experiments on radiation trapping \cite{Labeyrie:2003_,Labeyrie:2005_}, where a small beam at the center of the atomic cloud was used. That is why the relative amplitude of the slow decay is low even on resonance.

\subsection{Spectrum of the probe laser}

In our experiment, we used a distributed-feedback (DFB) laser diode as probe. This laser is expected to have a spectral linewidth on the order of $\sim 3$~MHz \cite{Kraft:2005_} corresponding to $0.5 \Gamma$, which should not have any influence at large detuning. However, if the spectrum has slow-decaying wings (e.g., Lorentzian), the amount of light a few $\Gamma$'s from the central frequency might be not negligible in the experiment.

\begin{figure}[b]
\centering
\includegraphics{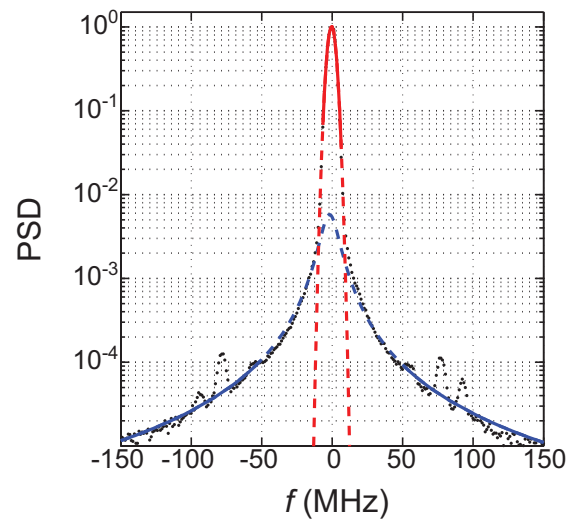}
\caption{Spectrum of the probe laser. The power spectrum density (PSD) of a beat-note signal with a reference laser has been averaged 100 times (scanning time $\sim 1$~s) and the central frequency ($\sim 1$~GHz) has been shifted to zero (black dots). The result can be fitted by a Gaussian in the central part (red) and by a Lorentzian in the wings (blue).}
\label{FigS4}
\end{figure}

\begin{figure*}
\centering
\includegraphics{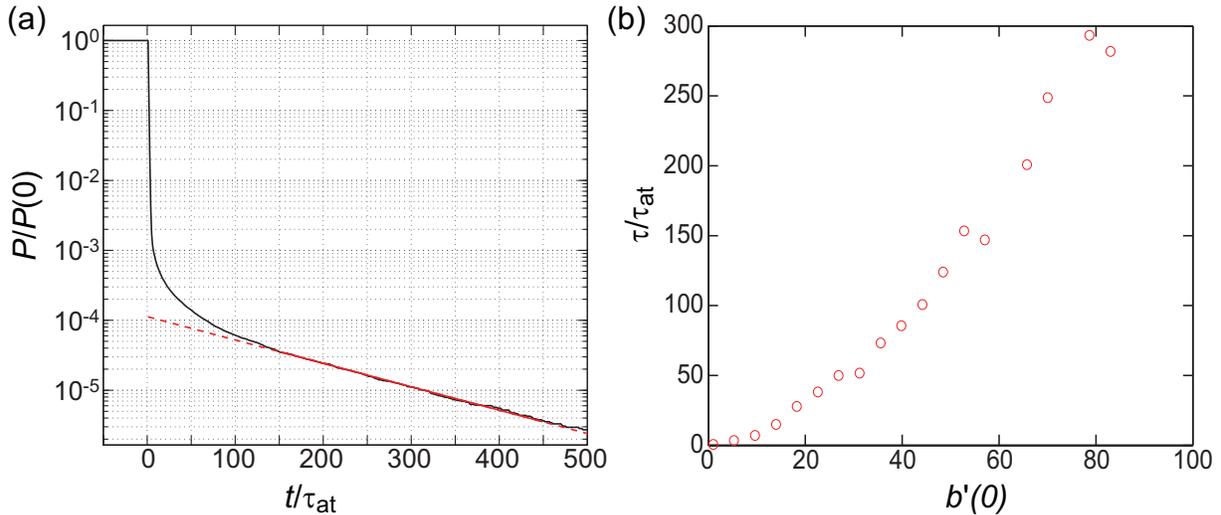}
\caption{Main results of a random walk model. (a) Temporal decay computed with $b'(0) = 53$ and $\delta = -6$. The relative amplitude of the slow part is $10^{-4}$. (b) Fitted time constant of the slow tail as a function of the on-resonance optical thickness. The increase is clearly faster than linear.}
\label{FigS5}
\end{figure*}

To characterize the spectrum of our probe laser, we performed a beat-note experiment with a commercial extended-cavity laser diode (Toptica DL pro), with a much narrower spectral width. It is thus a good (but slightly conservative) approximation to consider the power spectrum density of the beat-note signal, recorded with an electronic spectrum analyzer, as the optical spectrum of our probe laser. Such a measurement is shown in Fig.~\ref{FigS4}. The spectrum is composed of a central part, which can be well fitted by a Gaussian, superimposed on a small component with large wings, which is well fitted by a Lorentzian. The r.m.s. width of the Gaussian is $2.6$~MHz (corresponding to a full width at half maximum of $1 \Gamma$, larger than expected) and the Lorentzian wings have a relative amplitude of $5.8\times 10^{-3}$ and a full width at half maximum $\gamma \approx 13.2$~MHz corresponding to $2.2 \Gamma$.

From the measured spectrum, we can numerically compute the amount of light near resonance (in a width $\Gamma$) given the detuning of the central frequency. For example, for $\delta = -6 \Gamma$, corresponding to the largest detuning used in the measurements, this relative amount is $\approx 1.5 \times 10^{-4}$. We recall that even when the laser is set on resonance, the relative amplitude of the slow decay is only $A_\sub \simeq 2\times 10^{-2}$ so that the two numbers should be multiplied to predict the relative amplitude of the slow decay induced by the small part of resonant light. This is thus much too small to be visible in the experiment.

\subsection{Random walk model with frequency redistribution}

One further spurious effect is frequency redistribution induced by the Doppler effect during the multiple scattering process, which is known to play an important role in radiation trapping experiments in cold atoms~\cite{Labeyrie:2003_,Labeyrie:2005_,Pierrat:2009_}. First, for light initially on-resonance, the light spectrum gets broader during the multiple scattering and the effective optical thickness decreases, breaking the $\tau \propto b^2$ scaling for high enough temperature.
Second, light initially slightly detuned has a tendency to drift towards resonance (in addition to the broadening). It means that the previous evaluation considering a window of width $\Gamma$ to define the resonant light may be not appropriate because frequency redistribution induces a broader `capture range' of frequency.

To evaluate these effects, we performed numerical simulations considering the random walk of photons in a homogeneous spherical cloud of atoms, taking into account the Doppler effect and its associated frequency redistribution.
We send photons one by one, we randomly draw their initial transverse positions and detuning taking into account the spatial shape and measured frequency spectrum of the probe laser, and we compute how many scattering events the photons undergo before escaping, which is converted to a time, using the fact that each scattering event takes in average $\tau_\at$~\cite{Labeyrie:2003_}. We obtain thus a distribution of escape time, which is then convoluted by the duration of the pulse in order to simulate the fluorescence decay at the switch-off.

In Fig.~\ref{FigS5}(a), we show the fluorescence decay for a laser detuning centered at $\delta = -6$, a cloud with a temperature $T=50~\mu$K and an optical thickness $b'(0) = 53$, corresponding to $b_0 \simeq 110$ [Eq.~(1)], for comparison with the data of Figs. 2-3 of the main paper. We see indeed that the small amount of light that is on resonance induces a slow decay, whose relative amplitude is $\sim 10^{-4}$. As anticipated, this is larger than expected from the previous simple evaluation, but it is still more than one order of magnitude smaller than the subradiance decay observed in the experiment. Moreover, the use of a homogeneous sphere instead of a Gaussian density distribution overestimates substantially this amplitude because for a given peak-optical thickness (as measured by the transmission of a small beam at the center of the cloud), the average optical thickness `seen' by a large beam is larger with a homogeneous sphere than with a Gaussian cloud. We have also checked that if we consider an incident short pulse of duration similar to the switch-off of the probe beam, and its associated spectral broadening, the resulting slow decay also has a relative amplitude below $10^{-4}$.

Finally, we show in Fig.~\ref{FigS5}(b) how the time constant associated with this small, slow decay, which can be fitted by a decaying exponential, evolves with the optical depth of the cloud. We can see that the increase with the optical depth is much faster than linear in this regime of low temperature~\cite{Pierrat:2009_}. This provides another, qualitative difference between our subradiance data and what could be expected from spurious radiation trapping of near-resonant light.


\end{document}